\documentclass{aastex61}
\usepackage{amsmath}
\usepackage{float}
\usepackage{graphicx}
\usepackage{listings}
\usepackage{rotating}
\usepackage{longtable}

\newcommand{\lapprox} {\, \lower3pt\hbox{$\sim$}\llap{\raise2pt\hbox{$<$}}\,}
\newcommand{\gapprox} {\, \lower3pt\hbox{$\sim$}\llap{\raise2pt\hbox{$>$}}\,}

\begin{document}

\title{Coronal Loop Scaling Laws for Various Forms of Parallel Heat Conduction}
\correspondingauthor{Stephen J. Bradshaw}
\email{stephen.bradshaw@rice.edu}

\author[0000-0002-3300-6041]{Stephen J. Bradshaw}
\affiliation{Department of Physics \& Astronomy, Rice University, Houston, TX 77005, USA}

\author[0000-0001-8720-0723]{A. Gordon Emslie}
\affiliation{Department of Physics \& Astronomy, Western Kentucky University, Bowling Green, KY 42101, USA}

\author{Nicolas H. Bian}
\affiliation{Department of Physics \& Astronomy, Western Kentucky University, Bowling Green, KY 42101, USA}

\author[0000-0002-8078-0902]{Eduard P. Kontar}
\affiliation{School of Physics \& Astronomy, University of Glasgow, Glasgow G12 8QQ, Scotland, UK}

\begin{abstract}
The solar atmosphere is dominated by loops of magnetic flux which connect the multi-million-degree corona to the much cooler chromosphere. The temperature and density structure of quasi-static loops is determined by the continuous flow of energy from the hot corona to the lower solar atmosphere. Loop scaling laws provide relationships between global properties of the loop (such as peak temperature, pressure, and length); they follow from the physical variable dependencies of various terms in the energy equation, and hence the form of the loop scaling law provides insight into the key physics that controls the loop structure. Traditionally, scaling laws have been derived under the assumption of collision-dominated thermal conduction. Here we examine the impact of different regimes of thermal conduction -- collision-dominated, turbulence-dominated, and free-streaming -- on the form of the scaling laws relating the loop temperature and heating rate to its pressure and half-length. We show that the scaling laws for turbulence-dominated conduction are fundamentally different than those for collision-dominated and free-streaming conduction, inasmuch as the form of the scaling laws now depend primarily on conditions at the low-temperature, rather than high-temperature, part of the loop. We also establish regimes in temperature and density space in which each of the applicable scaling laws prevail.
\end{abstract}


\section{Introduction}
\label{introduction}
In a landmark paper based on {\it Skylab} EUV observations, \cite{1978ApJ...220..643R} showed that the solar corona is not a plane-parallel structure, but rather is dominated by approximately isobaric loop-like structures with shapes controlled by the structure of the ambient magnetic field. The plasma confined by the magnetic loops is heated to multi-million degree temperatures in the corona (by a mechanism that remains largely an open question), while the energy balance in the cooler lower atmosphere is dominated by radiative losses that dissipate the energy carried downward by thermal conduction. Hence, in quasi-static equilibrium, the plasma properties of coronal loops are strongly dependent on the conductive flux, which is usually taken to be proportional to the field-aligned temperature gradient \cite[see][for reviews]{2004psci.book.....A,2014LRSP...11....4R}.

\cite{1978ApJ...220..643R} studied the energy balance between heating, radiation, and field-aligned thermal conduction in quasi-static loops to deduce the now-well-known ``scaling laws'' relating the maximum temperature $T_M$ (K) and heating rate $E_H$ (erg~cm$^{-3}$~s$^{-1}$) in the loop to its half-length $L$ (cm) and pressure $p$ (dyne~cm$^{-2}$), viz.

\begin{equation}\label{rtv43}
T_M \simeq 1.4 \times 10^3 \, (pL)^{1/3}
\end{equation}
(their Equation~(4.3)) and

\begin{equation}\label{rtv44}
E_H \simeq 9.8 \times 10^4 \, p^{7/6} \, L^{-5/6}
\end{equation}
(their Equation~(4.4)). \cite{1995ApJ...454..934K} found a different coefficient and index for the temperature scaling law ($3.8\times10^4$ and $1/(5.1\pm0.5)$, respectively), based on observational measurements and line-fitting, and they discuss possible reasons for these discrepancies. The modeling in \cite{1978ApJ...220..643R} assumed uniform volumetric heating \citep[a condition later relaxed by][]{1981ApJ...243..288S,2010ApJ...714.1290M}, optically thin radiative losses, and heat conduction dominated by collisional transport of electrons, in which the heat flux is proportional to the local temperature gradient, with a coefficient that is temperature-dependent. However, in certain situations (that we discuss in Section~\ref{regimes}) the heat flux can become saturated (and hence a function of only local density and temperature conditions, rather than the temperature gradient) and reaches its free-streaming limit \citep{1975PhFl...18.1299M,1977ApJ...211..135C,1984PhRvA..30..365C}, or the mean free path used in calculating the thermal conduction coefficient may be limited by some form of turbulence.

Observations of coronal loop-top hard X-ray sources in solar flares require that the bremsstrahlung-producing electrons are confined to the corona \citep[e.g.,][]{1994Natur.371..495M,1995ApJ...440..370D,1996ApJ...459..815M,1997ApJ...478..787T,2004ApJ...603L.117V,2008A&ARv..16..155K,2008PASJ...60..835J,2012ApJ...755...32G,2013A&A...551A.135S}. Various authors have considered mechanisms that could be responsible for more effective confinement of accelerated electrons in the corona, and in particular the possibility that turbulence enhances the angular scattering rate and so suppresses the rate of escape of \emph{non-thermal} electrons from the coronal acceleration region \citep[e.g.,][]{2014ApJ...780..176K,2017ApJ...835..262B}. Of course, the presence of such turbulence will also act to suppress energy transport by {\it thermal} electrons. Further, it is likely that some form of small-scale turbulence exists in active region loops, particularly if coronal heating is due to small-scale processes such as the flux-braiding and reconnection mechanism described by \cite{1988ApJ...330..474P} or the interaction of counter-propagating Alfv\'{e}n waves \citep{2011ApJ...736....3V}. It is therefore of interest to consider the effect of small-scale turbulence on the thermal conductive flux and hence on the form of the loop scaling laws. At the other extreme, if the scattering length (either collisional or turbulent) is large compared to the loop half-length $L$, then the heat flux can become saturated.

In this paper we consider how modifications to the collision-dominated physics of thermal conduction, either toward suppression of conductive flux by turbulence or toward the free-streaming limit, affect the form of the scaling laws appropriate to active region loops.  Our expressions modify those of \cite{1978ApJ...220..643R} in not only a quantitative but, as we shall see, qualitative, way, in which conditions in the transition region of the loop can play a role comparable to, or even more important than, conditions in the hot regions near the coronal apex. We shall also explore the physical conditions in which each of these regimes dominates the form of the conductive flux and hence the form of the pertinent scaling law.

We begin in Section~\ref{analysis} with a review of the fundamental energy balance equation in a static coronal loop structure, and we consider the form of this equation in regimes for which energy transport by thermal conduction takes place by a combination of collisional, turbulent, or free-streaming processes. Using some plausible assumptions, we derive approximate analytic solutions to this energy equation and thus deduce the corresponding loop scaling laws for both peak temperature and heating rate in terms of the pressure in, and length of, the loop structure.  In Section~\ref{regimes} we discuss the physical regimes in which each of these scaling laws apply, and in Section~\ref{discussion} we discuss and summarize our findings.

\section{Derivation of the Coronal Loop Scaling Laws for Various forms of Parallel Heat Conduction}
\label{analysis}

\subsection{Rigorous Scaling Laws}

Our starting point is the well-known energy equation for an isobaric coronal loop, which describes the balance between heat in, radiation out, and energy redistribution by electron-dominated thermal conduction along a guiding magnetic field line.  We delineate the loop using the one-dimensional coordinate $s$, measured from the loop base toward the loop apex, i.e., in the direction of positive temperature gradient and in general antiparallel to the direction of the heat flux.  The quasi-static energy balance is thus given by

\begin{equation}\label{energy_equation}
\frac{dF}{ds} + E_R = E_H \,\,\, ,
\end{equation}
where $E_H$ (erg~cm$^{-3}$~s$^{-1}$) is the heat input,

\begin{equation}\label{radiative_losses}
E_R = n^2 \Lambda(T) = \frac{p^2}{4 k_B^2 T^2} \, \Lambda(T)
\end{equation}
is the (optically thin) radiative loss (erg~cm$^{-3}$~s$^{-1}$) and $F$ (erg~cm$^{-2}$~s$^{-1}$) is the conductive flux (aligned in a direction antiparallel to the temperature gradient). For thermal transport dominated by Coulomb collisions (denoted below by the notation $[C]$), there is the well-known \cite{1962pfig.book.....S} result

\begin{equation}\label{conduction-spitzer}
F_{C} = - \kappa_{o} \, T^{5/2} \, \frac{dT}{ds} \,\,\, . \qquad [C]
\end{equation}
In the above equations, $n$ (cm$^{-3}$) is the density, $T$ (K) is the electron temperature, $p=2n k_B T$ (erg~cm$^{-3}$) is the gas pressure, and

\begin{equation}\label{ko-def}
\kappa_{o} = \frac{k_B \, (2 k_B)^{5/2}}{\pi m_e^{1/2} e^4 \ln \Lambda} \simeq 1.7 \times 10^{-6} \, {\rm erg} \, {\rm cm}^{-1} \, {\rm s}^{-1} \, {\rm K}^{-7/2}
\end{equation}
is the \cite{1962pfig.book.....S} coefficient of thermal conductivity. $\Lambda(T)$ (erg~cm$^3$~s$^{-1}$) is the optically-thin radiative loss function which, in the temperature range of interest ($10^5$~K~$< T < 10^7$~K) is well approximated by the power-law form

\begin{equation}\label{rad-losses-power-law}
\Lambda(T) = \chi \, T^{-1/2} \,\,\, ,
\end{equation}
where $\chi \simeq 1.6 \times 10^{-19}$~erg~cm$^3$~s$^{-1}$~K$^{1/2}$.

As discussed by \cite{2016ApJ...824...78B}, the presence of (for example) a spectrum of magnetic field fluctuations within the loop gives rise to an additional source of angular scattering for electrons, hereafter referred to as ``turbulent scattering,'' with an associated (velocity-independent) mean free path $\lambda_T$ (cm).  For example, for turbulent scattering associated with local inhomogeneities $\delta B_\perp$ in a background magnetic field $B_0$,

\begin{equation}
\lambda_T = \lambda_B \, \left ( \frac{\delta B_\perp}{B_0} \right )^{-2} \,\,\, ,
\end{equation}
where $\lambda_B$ is the magnetic correlation length.  Although it is possible that the turbulent heat conductivity also depends on quantities such as the magnetic energy release rate (via the fluctuation energy $\delta B_{\perp}^{2}/8\pi$), we here, for simplicity, take $\lambda_{T}$ to be a constant parameter. When the heat flux is controlled by turbulent scattering, the expression for the heat flux becomes

\begin{equation}\label{conduction-turbulent}
F_{T} = - \frac{\kappa_{o}}{R} \, T^{5/2} \, \frac{dT}{ds} \,\,\, , \qquad [T]
\end{equation}
where we have introduced the notation $[T]$) and the turbulent heat flux correction factor $R$ \citep{2018ApJ...852..127B} reflects the ratio of the collisional to turbulent mean free paths:

\begin{equation}\label{R_expression}
R = \frac{\lambda_C}{\lambda_T} = \frac{(2 k_B T)^{2}}{ 2 \pi e^4 \, n \, \ln \Lambda \, \lambda_T} = \frac{(2 \, k_B \, T)^3}{2 \pi e^4 \, \ln \Lambda \, \lambda_T \, p} \equiv c_R \left ( \frac{T^3}{\lambda_T \, p} \right )\,\,\, ,
\end{equation}
where

\begin{equation}\label{cr-def}
c_R = \frac{4 k_B^3}{\pi e^4 \ln \Lambda} \simeq 3.15 \times 10^{-12}~{\rm erg} \, {\rm cm}^{-2}~{\rm K}^{-3} \,\,\, .
\end{equation}
Substituting Equation~({\ref{R_expression}) in Equation~(\ref{conduction-turbulent}) gives

\begin{equation}\label{conduction-turbulent-revised}
F_{T} = - \kappa_{o} \, \left ( \frac{\lambda_T \, p}{c_R} \right ) \, T^{-1/2} \, \frac{dT}{ds} \,\,\, . \qquad [T]
\end{equation}

When the mean free path (either collisional or turbulent) becomes larger than the characteristic scale of the loop (e.g., its half-length $L$), the thermal conductive flux is no longer inhibited by scattering processes and so approaches its free-streaming value.  Since the electrons at a given point now originate from a wide range of positions (and so temperatures) within the loop, the value of the thermal conductive flux at a given point is in general a non-local quantity, formed by the convolution of the expression for the local conductive flux as a function of temperature $T$ with the temperature profile of the loop \citep{2018ApJ...865...67E}. Since we are here interested in 0-D global scaling laws, rather than 1-D variations in quantities with position, we will neglect this non-local factor, which by construction averages to zero over the loop.  An upper limit to the heat flux is therefore set by the free-streaming limit in which particles move in the direction antiparallel to the local temperature gradient at the local thermal speed \citep[see discussion in][]{Bradshaw2006}:

\begin{equation}\label{max_heat_flux}
F_{\rm max} = - E_{th} \, v_{th} \,\,\, ,
\end{equation}
where $E_{th}= (3/2) \, n k_B T$ is the electron thermal energy density and $v_{th} = \sqrt{k_B T/ m_e}$ is the thermal speed. A correcting factor is usually also employed and generally, based on Fokker-Planck simulations \citep[e.g.,][]{2008ApJ...682.1351K}, is taken to be 1/6. The maximum electron heat flux (denoted by $[S]$) is therefore

\begin{equation}\label{f-free-streaming}
F_{S} \simeq \frac{1}{6} \, \times \, \frac{3}{2} \, n k_B T \, \sqrt{\frac{k_B T}{m_e}} = \frac{1}{2^{7/2}} \, \frac{n (2 k_B T)^{3/2}}{m_e^{1/2}} \,\,\, . \qquad [S]
\end{equation}

We now proceed to derive the loop scaling laws that follow from the various expressions~(\ref{conduction-spitzer}), (\ref{conduction-turbulent}), and~(\ref{f-free-streaming}) for the conductive flux. For the collision-dominated and turbulence-limited cases, we start by recasting the energy equations in the form \citep[cf. Equation~(3.11) of][]{1978ApJ...220..643R}:

\begin{eqnarray}\label{flux-vs-temp}
\frac{F_{C}}{\kappa_{o} T^{5/2}} \, \frac{dF_{C}}{dT} & = & \frac{p^2}{4 k_B^2} \frac{\Lambda(T)}{T^2} - E_H \, ; \qquad [C]\cr
\frac{c_R T^{1/2} F_{T}}{\kappa_{o} \lambda_T p} \, \frac{dF_{T}}{dT} & = & \frac{p^2}{4 k_B^2} \frac{\Lambda(T)}{T^2} - E_H \,\,\, . \qquad [T]
\end{eqnarray}
Using Equation~(\ref{rad-losses-power-law}),  Equations~(\ref{flux-vs-temp}) may be written

\begin{eqnarray}\label{fc-eqn}
F_{C} \, dF_{C}  & = &  \kappa_{o} \left [ \frac{\chi p^2}{4 k_B^2} \, dT - E_H T^{5/2} \, dT \right ] \, ; \qquad [C] \cr
F_{T} \, dF_{T}  & = &  \frac{\kappa_{o} \, \lambda_T \, p}{c_R} \left [ \frac{\chi p^2}{4 k_B^2} \, \frac{dT}{T^3} - E_H \frac{dT}{T^{1/2}} \right ] \,\,\, , \qquad [T]
\end{eqnarray}
which can both be straightforwardly integrated from $T=T_{0}$ (the base of the transition region) to $T$ to give

\begin{eqnarray}\label{flux-integral}
F_{C}^2(T) - F_{C}^2(T_{0}) & = & \kappa_{o} \left [ \frac{\chi p^2}{2 k_B^2} \, ( T - T_{0}) - \frac{4 E_H}{7} \left ( T^{7/2} - T_{0}^{7/2} \right ) \right ] \,\,\, ; \qquad [C] \cr
F_{T}^2(T) - F_{T}^2(T_{0}) & = & \frac{\kappa_{o} \, \lambda_T \, p}{c_R} \left [ \frac{\chi p^2}{4 k_B^2} \, \left ( \frac{1}{T_{0}^2} -\frac{1}{T^2} \right ) - 4 E_H \left ( T^{1/2} - T_{0}^{1/2} \right ) \right ] \qquad [T]
\end{eqnarray}
\citep[the first of these can be compared with Equation~(3.12) of][]{1978ApJ...220..643R}.  In general \citep[see][]{1978ApJ...220..643R} the heat fluxes $F_{C}(T_{0})$ and $F_{T}(T_{0})$ at the lower boundary may be neglected.  We can then substitute for $F_{C}$ and $F_{T}$ from Equations~(\ref{conduction-spitzer}) and~(\ref{conduction-turbulent}), respectively, to obtain

\begin{eqnarray}\label{dtbds}
\left ( \frac{dT}{ds} \right )^2 & = & \frac{1}{\kappa_{o} T^5} \left [ \, \frac{\chi p^2}{2 k_B^2} (T - T_{0}) - \frac{4 E_H}{7} \left ( T^{7/2} - T_{0}^{7/2} \right ) \right ] \, ; \qquad [C] \cr
\left ( \frac{dT}{ds} \right )^2 & = & \frac{c_R T}{\kappa_{o} \lambda_T \, p} \, \left [ \frac{\chi p^2}{4 k_B^2} \left ( \frac{1}{T_{0}^2} - \frac{1}{T^2} \right ) - 4 E_H \left ( T^{1/2} - T_{0}^{1/2} \right ) \right ] \,\,\, . \qquad [T]
\end{eqnarray}
These can be integrated again to yield an expression for $s(T)$:

\begin{eqnarray}\label{sT}
s(T) - s(T_{0}) & = & \kappa_{o}^{1/2} \, \int_{T_{0}}^{T} \left [ \frac{\chi p^2}{2 k_B^2} ( T - T_{0}) - \frac{4 E_H}{7} \left ( T^{7/2} - T_{0}^{7/2} \right ) \right ]^{-1/2} \, T^{5/2} \, dT \,\,\, ; \qquad [C] \cr
s(T) - s(T_{0}) & = & \left ( \frac{\kappa_{o} \lambda_T p}{c_R} \right )^{1/2} \, \int_{T_{0}}^{T}  \left [ \frac{\chi p^2}{4 k_B^2} \left ( \frac{1}{T_{0}^2} - \frac{1}{T^2} \right ) - 4 E_H \left ( T^{1/2} - T_{0}^{1/2} \right ) \right ]^{-1/2} \, T^{-1/2} \, dT \,\,\, . \qquad [T]
\end{eqnarray}

At this juncture \cite{1978ApJ...220..643R} neglect the second term in the square brackets in the first of these equations, arguing that near the base of the loop the primary energy balance is between heat flux and transition region radiative losses.  We shall proceed somewhat differently here.  First, we note from Equation~(\ref{dtbds}) that since the temperature gradient vanishes at the loop apex ($T=T_{M}$),

\begin{eqnarray}\label{tm-condition}
\frac{\chi p^2}{2 k_B^2} (T_{M} - T_{0}) & = & \frac{4 E_H}{7} \left ( T_{M}^{7/2} - T_{0}^{7/2} \right ) \,\,\, ; \qquad [C] \cr
\frac{\chi p^2}{4 k_B^2} \left ( \frac{1}{T_{0}^2} - \frac{1}{T_{M}^2} \right ) & = & 4 E_H \left ( T_{M}^{1/2} - T_{0}^{1/2} \right ) \,\,\, . \qquad [T] \end{eqnarray}
Since $T_{M} \gg T_{0}$, Equation~(\ref{tm-condition}) gives, to a high degree of accuracy,

\begin{eqnarray}\label{tm-approx}
T_{M}^{5/2} \simeq \frac{7 \chi p^2}{8 k_B^2 E_H} \,\,\, ; \qquad [C] \cr
T_{M}^{1/2} \simeq \frac{\chi p^2}{16 k_B^2 E_H T_{0}^2} \,\,\, . \qquad [T]
\end{eqnarray}

It is important to note that in the case of collision-dominated conduction only positive powers of the temperature appear in Equation~(\ref{tm-condition}), and therefore the base electron temperature $T_{0}$ does not play a significant role in determining the maximum temperature $T_{M}$. Physically, this is because the temperature gradient $dT/ds \propto F_{C}/T^{5/2}$; thus the low temperatures at the loop base mean that the temperature gradient is very large (compared to that in the corona) in order to support the incident heat flux in the presence of a much lower thermal conduction coefficient.  Quantitatively, the thickness $\ell$ of a layer corresponding to a temperature range $T_{1} < T < T_{1} + \Delta T$ is given by $\Delta \ell = (dT/ds)^{-1} \, \Delta T \propto T^{5/2} \, \Delta T$, which is much smaller at the loop base ($T \simeq T_{0}$) than in the corona ($T \simeq T_{M}$).  Thus the amount of radiation emitted within such a layer is negligible and the loop energetics are controlled principally by a balance between heat conduction and radiation in the corona.

On the other hand, for turbulence-limited conduction the heat flux is reduced by a temperature-dependent factor $R \propto T^3$ (Equation~(\ref{R_expression})), so that the conductive coefficient in the expression for $F_T$ is {\it inversely} proportional to $T$: $F_T \propto T^{-1/2} dT/ds$ (Equation~(\ref{conduction-turbulent-revised})).  Thus the temperature gradient $dT/ds \propto F_{T} \, T^{1/2}$ (cf. Equations~(\ref{conduction-turbulent}) and~(\ref{R_expression})) and the thickness of a layer corresponding to a temperature difference $\Delta T$ is $\Delta \ell \propto T^{-1/2} \, \Delta T$, a quantity that is now {\it larger} at the loop base than in the corona (the temperature gradient need no longer steepen to support the incident heat flux because it is turbulence-limited).  As a result of this fundamentally different scaling of temperature gradient with temperature, {\it radiation from the low-temperature material near the base of the loop becomes more important than radiation from the corona}.  This effect is sufficiently strong that the integral in Equation~(\ref{sT}) becomes dominated by the low-temperature, rather than the high-temperature, domain and, as a result, {\it conditions at the loop base control the scaling laws for the loop}.

Using Equation~(\ref{tm-condition}) in Equation~(\ref{sT}), we find that the loop half-length $L = s(T_{M}) - s(T_{0})$ is given by

\begin{eqnarray}\label{L-approx}
L & = & \left ( \frac{7\kappa_{o}}{4 E_H} \right )^{1/2} \, \int_{T_{0}}^{T_{M}}  \left [ \left ( \frac{T_{M}^{7/2} - T_{0}^{7/2}}{T_{M} - T_{0}} \right ) (T - T_{0}) - \left ( T^{7/2} - T_{0}^{7/2} \right ) \right ]^{-1/2} \, T^{5/2} \, dT  \cr
& \simeq & \left ( \frac{7\kappa_{o}}{4 E_H} \right )^{1/2} \, \int_{T_{0}}^{T_{M}}  \left [ T_{M}^{5/2} \, T - T^{7/2} \right ]^{-1/2} \, T^{5/2} \, dT  \cr
& \simeq & \left ( \frac{7\kappa_{o}}{4 E_H} \right )^{1/2} \, \int_{T_{0}}^{T_{M}} \frac{T^2 \, dT}{\left ( T_{M}^{5/2} - T^{5/2} \right )^{1/2}} =
\frac{2}{5} \, \left ( \frac{7\kappa_{o}}{4 E_H} \right )^{1/2} \, \int_0^{T_{M}^{5/2}-T_{0}^{5/2}} (T_{M}^{5/2} - x)^{1/5} \frac{dx}{x^{1/2}} \cr
& \simeq & \frac{2}{5} \left ( \frac{7\kappa_{o}}{4 E_H} \right )^{1/2} \, T_{M}^{1/2} \int_0^{T_{M}^{5/2}-T_{0}^{5/2}} \frac{dx}{x^{1/2}} \simeq \frac{4}{5} \, \left ( \frac{7\kappa_{o}}{4 E_H} \right )^{1/2} \,  T_{M}^{7/4} = \left ( \frac{28\kappa_{o}}{25 E_H} \right )^{1/2} \, T_{M}^{7/4} \,\,\, ; \qquad [C] \cr
L & = & \left ( \frac{\kappa_{o} \lambda_T p}{4 c_R E_H} \right )^{1/2} \, \int_{T_{0}}^{T_{M}} \left [ \frac{ T_{M}^{1/2} - T_{0}^{1/2}}{ \left ( 1 - \frac{T_{0}^2}{T_{M}^2} \right )} \left (1 - \frac{T_{0}^2}{T^2} \right ) - \left ( T^{1/2} - T_{0}^{1/2} \right ) \right ]^{-1/2} \frac{dT}{T^{1/2}} \cr
& \simeq & \left ( \frac{\kappa_{o} \lambda_T p}{4 c_R E_H} \right )^{1/2} \, \int_{T_{0}}^{T_{M}} \left [ T_{M}^{1/2} \left (1 - \frac{T_{0}^2}{T^2} \right ) - \left ( T^{1/2} - T_{0}^{1/2} \right ) \right ]^{-1/2} \frac{dT}{T^{1/2}} \cr
& \simeq & \left ( \frac{\kappa_{o} \lambda_T p}{4 c_R E_H} \right )^{1/2} \, \int_{T_{0}}^{T_{M}}  \left [ T_{M}^{1/2} -  T^{1/2} \right ]^{-1/2} \, \frac{dT}{T^{1/2}} \cr
& = & \left ( \frac{\kappa_{o} \lambda_T P}{c_R E_H} \right )^{1/2} \, \int_{0}^{T_{M}^{1/2} - T_{0}^{1/2}} x^{-1/2} dx = \left ( \frac{4 \kappa_{o} \lambda_T p}{c_R E_H} \right )^{1/2} \, \left ( T_{M}^{1/2} - T^{1/2} \right )^{1/2} \left \vert_{T_{M}}^{T_{0}} \right . \simeq \left ( \frac{4 \kappa_{o} \lambda_T p}{c_R E_H} \right )^{1/2} T_{M}^{1/4} \,\,\, . \quad [T]
\end{eqnarray}
Eliminating $E_H$ between Equations~(\ref{tm-approx}) and~(\ref{L-approx}) results in the scaling laws

\begin{eqnarray}\label{scaling-law-tpl}
T_{M} & = & \left ( \frac{25 \chi}{32 \kappa_{o} k_B^2} \right )^{1/6} \, (pL)^{1/3} \, \simeq 1.3 \times 10^3 \, (pL)^{1/3} \,\,\, ; \qquad [C] \cr
T_{M} & = & \frac{\chi c_R}{64 \kappa_{o}  \lambda_T (k_B T_{0})^2} \, p L^2 \, \simeq \frac{2.4 \times 10^5}{\lambda_T \, T_{e0}^2} \, p L^2 \,\,\, . \qquad [T] \end{eqnarray}
Also, eliminating $T_{M}$ between Equations~(\ref{tm-approx}) and~(\ref{L-approx}) gives

\begin{eqnarray}\label{scaling-law-ehpl}
E_H & = & \frac{28}{25} \, \kappa_{o} \, \left ( \frac{25 \chi}{32 \, \kappa_{o} \, k_B^2}  \right )^{7/12} \, p^{7/6} \, L^{-5/6} \, \simeq 1.3 \times 10^5 \, p^{7/6} \, L^{-5/6} \,\,\, ; \qquad [C] \cr
E_H & = & \left ( \frac{\kappa_{o} \chi \lambda_T}{4 \, c_R}  \right )^{1/2} \, \left ( \frac{1}{k_B T_{0}} \right ) \, p^{3/2} \, L^{-1} \, \simeq \frac{1.1 \times 10^9 \, \lambda_T^{1/2}}{T_{0}} \, p^{3/2} \, L^{-1} \,\,\, , \qquad [T]
\end{eqnarray}
where we have substituted values for $\chi$, $\kappa_o$ and $k_B$.  The first line in each of the results in Equations~(\ref{scaling-law-tpl}) and~(\ref{scaling-law-ehpl}) are the \cite{1978ApJ...220..643R} results expressed in Equations~(\ref{rtv43}) and~(\ref{rtv44}) above, with slightly different coefficients because of the different value of the Coulomb logarithm $\ln \Lambda$, and hence $\kappa_o$, used.

The expressions for the case of free-streaming heat flux are developed somewhat differently \citep[see also][who first applied thermal conduction in the free-streaming limit to coronal loop models]{1993SoPh..145...45C}. Using expressions~(\ref{radiative_losses}) and~(\ref{f-free-streaming}) in the basic energy equation~(\ref{energy_equation}), we find that

\begin{equation}\label{energy-equation-streaming}
\left(\frac{1}{2^{7/2}} \, \frac{n (2 k_B)^{3/2}}{m_e^{1/2}} \right)\, \, \frac{3}{2} \, T^{1/2} \, \frac{dT}{ds} + E_H = \frac{p^2}{4 k_B^2 T^2} \, \chi \, T^{-1/2} \,\,\, ,
\end{equation}
from which

\begin{equation}\label{dtbds-free}
\frac{dT}{ds} = \frac{2^{9/2}}{3} \, \frac{m_e^{1/2}}{n (2 k_B)^{3/2}} \, \frac{1}{T^{1/2}} \, \left [ \frac{p^2 \chi}{4 k_B^2 T^{5/2}} - E_H \right ] \,\,\, .
\end{equation}
Setting $dT/ds = 0$ at the apex ($T = T_{M}$) gives

\begin{equation}\label{eh-streaming}
E_H = \frac{p^2 \chi}{4 k_B^2 T_{M}^{5/2}} \,\,\, . \qquad [S]
\end{equation}
Substituting this in Equation~(\ref{dtbds-free}), using the relation $n = p/2k_B T$, inverting to get an expression for $ds/dT$, and integrating this expression from the base temperature $T_{0}$ to the apex temperature $T_{M}$ gives

\begin{equation}\label{st-free}
pL = \frac{3 \, (2 k_B)^{5/2}}{2^{9/2} m_e^{1/2} \, \chi} \, \int_{T_{0}}^{T_{M}} \frac{T^2 \, dT}{1 - ( T/T_{M} )^{5/2}} \simeq \frac{(2 k_B)^{5/2} \, T_{M}^3}{2^{9/2} m_e^{1/2} \, \chi} \,\,\, .
\end{equation}
This gives the scaling law

\begin{equation}\label{scaling-law-tp3}
T_{M} = \left ( \frac{4 m_e^{1/2} \chi}{k_B^{5/2}} \right )^{1/3} (pL)^{1/3} \simeq 4.4 \times 10^2 \, (pL)^{1/3} \,\,\, , \qquad [S]
\end{equation}
and using this in Equation~(\ref{eh-streaming}) gives the additional scaling law

\begin{equation}\label{scaling-law-ehp3}
E_H = \frac{\chi}{4 k_B^2} \, \left ( \frac{k_B^{5/2}}{4 m_e^{1/2} \chi} \right )^{5/6} \, p^{7/6} \, L^{-5/6} \simeq 5.1 \times 10^5 \, p^{7/6} \, L^{-5/6} \,\,\, . \qquad [S]
\end{equation}

To summarize, for electron-dominated conduction we have the following scaling laws in the three cases (collisional, turbulent, and free-streaming):

\begin{eqnarray}\label{scaling-law-tp-all}
T_{M} & = & \left ( \frac{25 \chi}{32 \kappa_{o} k_B^2} \right )^{1/6} \, (pL)^{1/3} \, \simeq 1.3 \times 10^3 \, (pL)^{1/3} \,\,\, ; \qquad [C] \cr
T_{M} & = & \frac{\chi c_R}{64 \kappa_{o}  \lambda_T (k_B T_{0})^2} \, p L^2 \, \simeq \frac{2.4 \times 10^5}{\lambda_T \, T_{0}^2} \, p L^2 \,\,\, ; \qquad [T] \cr
T_{M} & = & \left ( \frac{4 m_e^{1/2} \chi}{k_B^{5/2}} \right )^{1/3} (pL)^{1/3} \simeq 4.4 \times 10^2 \, (pL)^{1/3} \,\,\, ; \qquad [S]
\end{eqnarray}
and

\begin{eqnarray}\label{scaling-law-ehp-all}
E_H & = & \frac{28}{25} \, \kappa_{o} \, \left ( \frac{25 \chi}{32 \, \kappa_{o} \, k_B^2}  \right )^{7/12} \, p^{7/6} \, L^{-5/6} \, \simeq 1.3 \times 10^5 \, p^{7/6} \, L^{-5/6} \,\,\, ; \qquad [C] \cr
E_H & = & \left ( \frac{\kappa_{o} \chi \lambda_T}{4 \, c_R}  \right )^{1/2} \, \left ( \frac{1}{k_B T_{0}} \right ) \, p^{3/2} \, L^{-1} \, \simeq \frac{1.1 \times 10^9 \, \lambda_T^{1/2}}{T_{0}} \, p^{3/2} \, L^{-1} \,\,\, ; \qquad [T] \cr
E_H & = & \frac{\chi}{4 k_B^2} \, \left ( \frac{k_B^{5/2}}{4 m_e^{1/2} \chi} \right )^{5/6} \, p^{7/6} \, L^{-5/6} \simeq 5.1 \times 10^5 \, p^{7/6} \, L^{-5/6} \,\,\, . \qquad [S]
\end{eqnarray}

It should be noted that even though the physics is substantially different, the dependencies of both $T_M$ and $E_H$ on $p$ and $L$ are identical for the collisional ($[C]$) and free-streaming ($[S]$) scaling laws (even though their coefficients are a factor of 3-4 different).

\subsection{Approximate Scaling Laws}

We have seen above that for collision-dominated conduction the low temperature plasma can be neglected.  Indeed, if we ignore the low-temperature plasma in the loop and simply equate the magnitudes of the conductive and radiative heating terms at the location of peak temperature, we obtain

\begin{equation}\label{naive-comparison-classical}
\frac{2}{7} \, \kappa_{o} \frac{T_{M}^{7/2}}{L^2} = \frac{p^2}{4 k_B^2} \, \chi \, T_{M}^{-5/2} \,\,\, ,
\end{equation}
which gives

\begin{equation}\label{naive-scaling-law-1}
T_{M} = \left ( \frac{7\chi}{8 \kappa_{o} k_B^2 } \right )^{1/6} \, (pL)^{1/3} \,\,\, . \qquad [C]
\end{equation}
This differs from the more exact scaling law in Equation~(\ref{scaling-law-tp-all}) by the factor $[(7/8) \times (32/25)]^{1/6} \simeq 1.02$, within 2\% of unity and well within the approximations used in, for example, the value of the Coulomb logarithm or the assumption of a strict Maxwellian distribution.  Similarly, equating the heating term to the conduction term gives

\begin{equation}\label{naive-comparison-classical-2}
E_H = \frac{2}{7} \, \kappa_{o} \frac{T_{M}^{7/2}}{L^2} = \frac{2}{7} \, \kappa_{o} \left ( \frac{7 \chi}{8 \kappa_{o} k_B^2} \right )^{7/12} \, p^{7/6} L^{-5/6} \,\,\, , \qquad [C]
\end{equation}
where we have used Equation~(\ref{naive-scaling-law-1}). This differs from the more exact scaling law~(\ref{scaling-law-tp-all}) by the more substantial factor $(2/7) \times (25/28) \times [(7/8) \times (32/25)]^{7/12}$ $\simeq 0.27$.

Similarly, equating the loop-averaged divergence of the saturated conductive flux (Equation~(\ref{f-free-streaming})) with the radiative loss term at the peak (Equation~(\ref{rad-losses-power-law})) gives

\begin{equation}\label{naive-comparison-saturated}
\frac{1}{2^{7/2}} \, p \, \left ( \frac{2 k_B}{m_e} \right )^{1/2} \, \frac{T_{M}^{1/2}}{L} = \frac{p^2}{4 k_B^2} \, \chi \, T_{M}^{-5/2} \,\,\, .
\end{equation}
From this we find

\begin{equation}\label{naive-scaling-law-1-sat}
T_{M}  =  \left ( \frac{2 m_e^{1/2} \chi}{k_B^{5/2}} \right )^{1/3} (pL)^{1/3} \,\,\, , \qquad [S]
\end{equation}
which differs from the more exact result (Equation~(\ref{scaling-law-tp-all})) by a factor of only $(1/2)^{1/3} \simeq 0.79$.  Then, equating the heating and conduction terms gives

\begin{equation}\label{naive-comparison-law-2-sat}
E_H = \frac{1}{2^{7/2}} \, p \, \left ( \frac{2 k_B}{m_e} \right )^{1/2} \, \frac{T_{M}^{1/2}}{L} = \frac{\chi}{4 k_B^2} \, \left ( \frac{k_B^{5/2}}{2 \, m_e^{1/2} \chi} \right )^{5/6} \, p^{7/6} \, L^{-5/6} \,\,\, , \qquad [S] \end{equation}
which differs from the more exact result (Equation~(\ref{scaling-law-ehp-all})) by a factor of $2^{5/6} \simeq 1.8$.

However, a similar simple exercise fails to determine the correct scaling laws for the case of turbulence-dominated conduction, because it neglects the important role that radiation from the lower transition region plays in determining the maximum loop temperature, and hence the conductive flux and the required heating rate to balance it.

\section{Regimes where each set of scaling laws applies}\label{regimes}

Using Equations~(\ref{conduction-spitzer}), (\ref{ko-def}), (\ref{conduction-turbulent}), (\ref{R_expression}), and~(\ref{f-free-streaming}), the ratios of the magnitudes of the heat fluxes in the various limits (turbulent, collisional, free-streaming) are

\begin{eqnarray}\label{conduction-comparison}
F_{T} : F_C : F_{S} & = & \frac{n (2 k_B)^{3/2}}{m_e^{1/2}} \, \lambda_T \, T^{1/2} \, \frac{dT}{ds} \,\,\, : \frac{(2 k_B)^{7/2}}{ 2 \pi e^4 \ln \Lambda \, m_e^{1/2}} \,  T^{5/2} \, \frac{dT}{ds} \, : \frac{1}{2^{7/2}} \, \frac{n (2 k_B T)^{3/2}}{m_e^{1/2}} \\
& \simeq & \frac{2}{3} \, \frac{n \, (2 k_B)^{3/2} \lambda_T}{m_e^{1/2}}\, \frac{T^{3/2}}{L} \,\,\, : \frac{2}{7} \, \frac{(2 k_B)^{7/2}}{ 2 \pi e^4 \ln \Lambda \, m_e^{1/2}} \,  \frac{T^{7/2}}{L} \, : \frac{1}{2^{7/2}} \, \frac{n (2 k_B)^{3/2} T^{3/2}}{m_e^{1/2}} \,\,\, ,
\end{eqnarray}
where $\ln \Lambda$ is the Coulomb logarithm. Since the conductive heat flux is proportional to the mean free path or length scale, 
the ratio of the heat fluxes is simply the ratio of the corresponding scale lengths:

\begin{equation}\label{length-comparison}
F_{T} : F_C : F_{max} \quad = \quad \lambda_T \, : \frac{3}{7} \, \lambda_C : \frac{3}{2^{9/2}} \, L \,\,\, ,
\end{equation}
where $\lambda_C$ is the collisional mean free path (Equation~(\ref{R_expression})).

In general, for a given set of physical conditions (temperature, density), the lowest of the three heat flux values (collision-dominated, turbulence-dominated, free-streaming) controls the flow of heat and hence determines the pertinent conductive flux regime and associated scaling law.  Since the heat flux is proportional to the corresponding length scale (Equation~(\ref{length-comparison})), the issue of selecting the pertinent conductive regime thus reduces to selecting which of the three length scales in Equation~(\ref{length-comparison}) is the smallest.
\begin{figure}
\centering 
\includegraphics[width=0.49\linewidth]{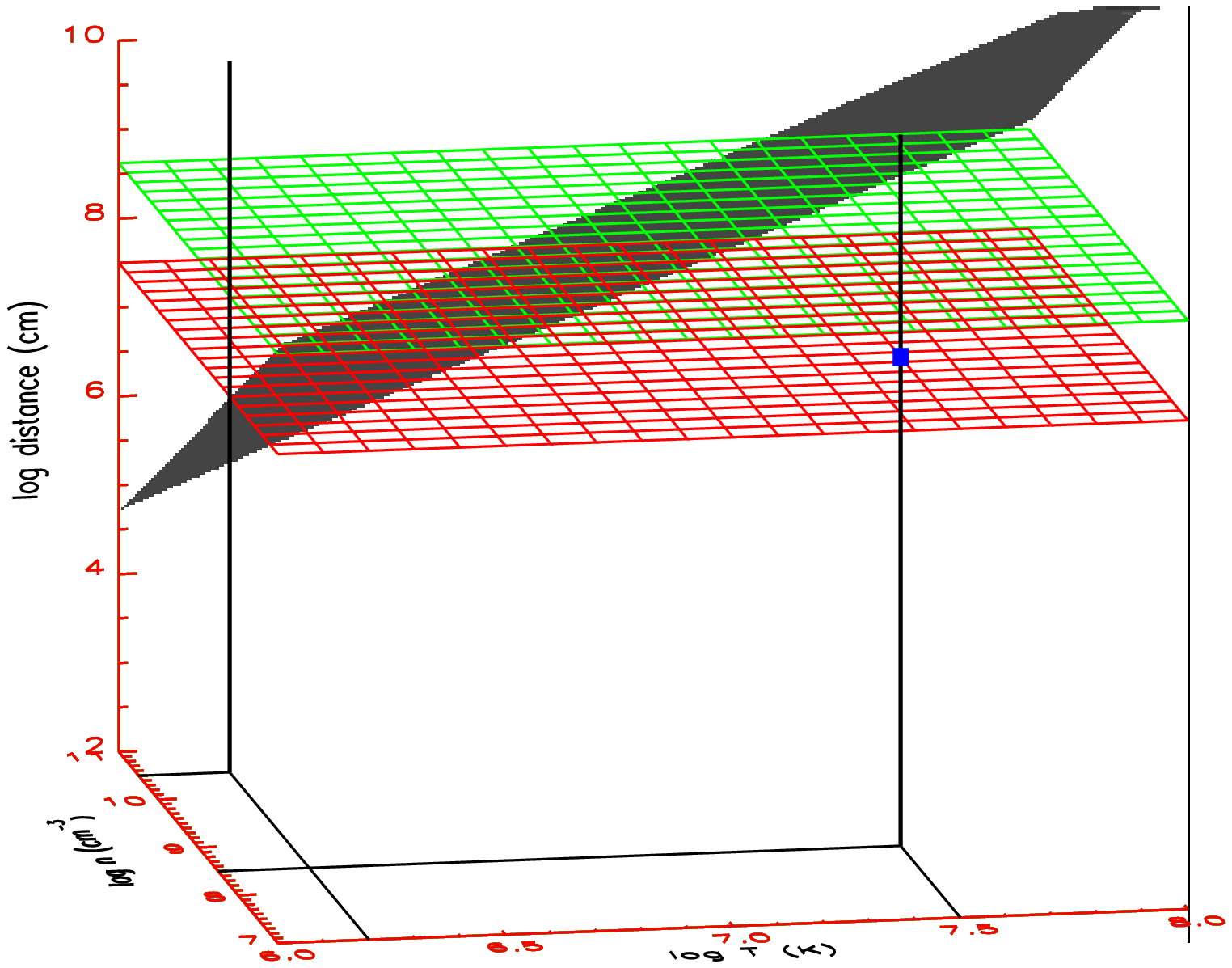}
\includegraphics[width=0.49\linewidth]{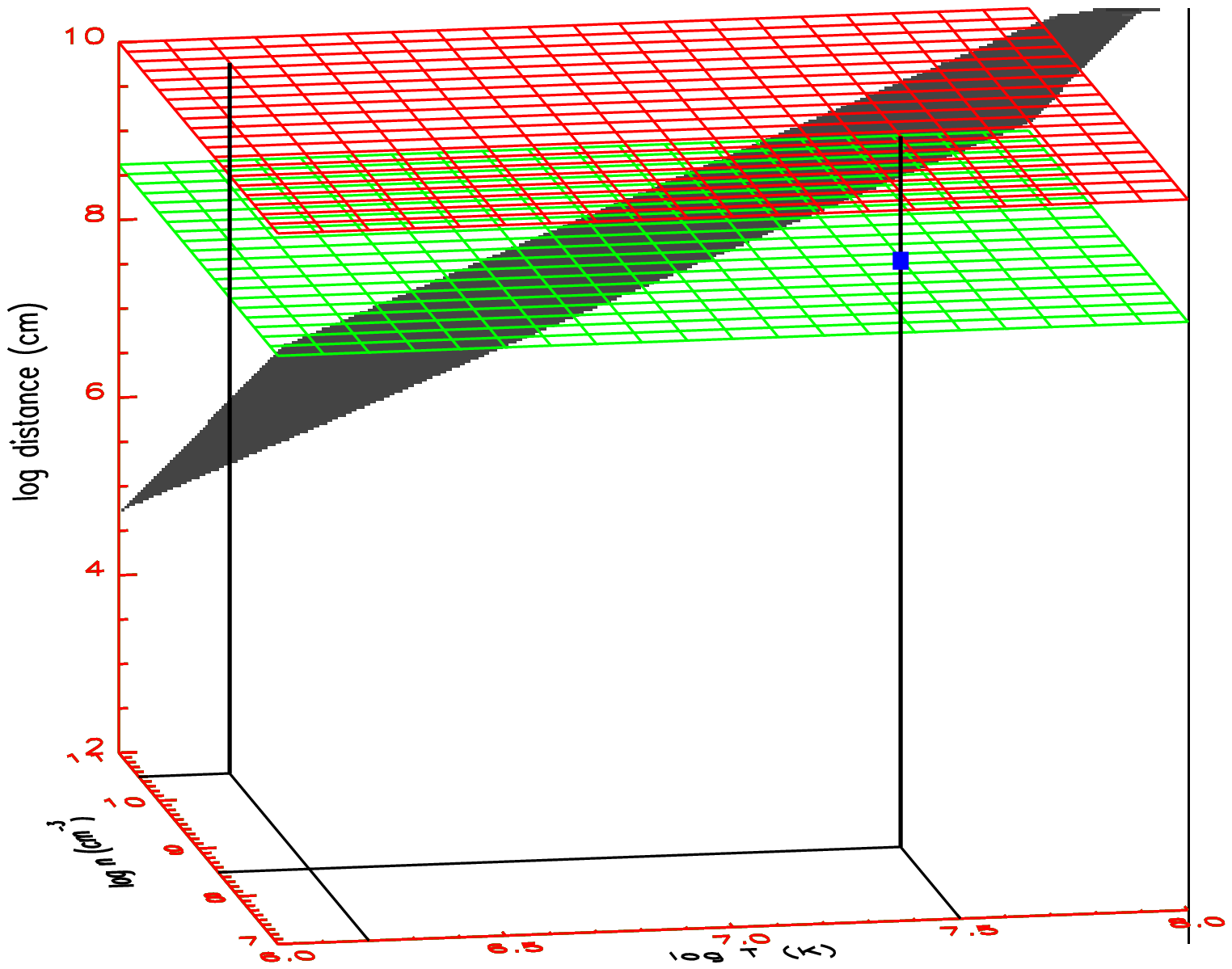}
\caption{Values of pertinent spatial scales for different conduction regimes. The slanted surface represents the quantity $(3/7) \, \lambda_C$, while the red and green horizontal surfaces represent the quantities $\lambda_T$ and $(3/2^{9/2}) \, L$, respectively (see Equation~(\ref{length-comparison})). The left figure corresponds to a turbulence scale $\lambda_T = 10^{7.5}$~cm \citep[cf.][]{2016ApJ...824...78B} and a loop half-length $L = 10^{9.5}$~cm ($3L/2^{9/2} \simeq 10^{8.6}$~cm); the right figure corresponds to a very long turbulence scale $\lambda_T = 10^{10}$~cm (and so weak turbulence) and the same loop half-length $L$. In both figures, the relative values of these three length scales are highlighted at two values of the plasma temperature $T$ (K) and density $n$ (cm$^{-3}$): ($T = 10^{6.2}, n = 10^{10.5}$) and ($T = 10^{7.5}, n = 10^{9.0}$).  For the latter set of $(n,T)$ values the blue dot marks where the vertical line crosses the horizontal surfaces denoting the values of $\lambda_T$ and $(3/2^{9/2}) \, L$, respectively.\label{scale-length-comparison}}
\end{figure}

Figure~\ref{scale-length-comparison} compares the values of the three length scales in Equation~(\ref{length-comparison}), for two cases: $(\lambda_T, L) = (10^{7.5}, 10^{9.5})$~cm (left panel), and $(\lambda_T, L) = (10^{10}, 10^{9.5})$~cm (right panel). Figure~\ref{scale-length-comparison-2D} shows a slice through each of the 3D plots in Figure~\ref{scale-length-comparison} at $T=10^7$~K by way of an example.
\begin{figure}
	\centering 
	\includegraphics[width=0.49\linewidth]{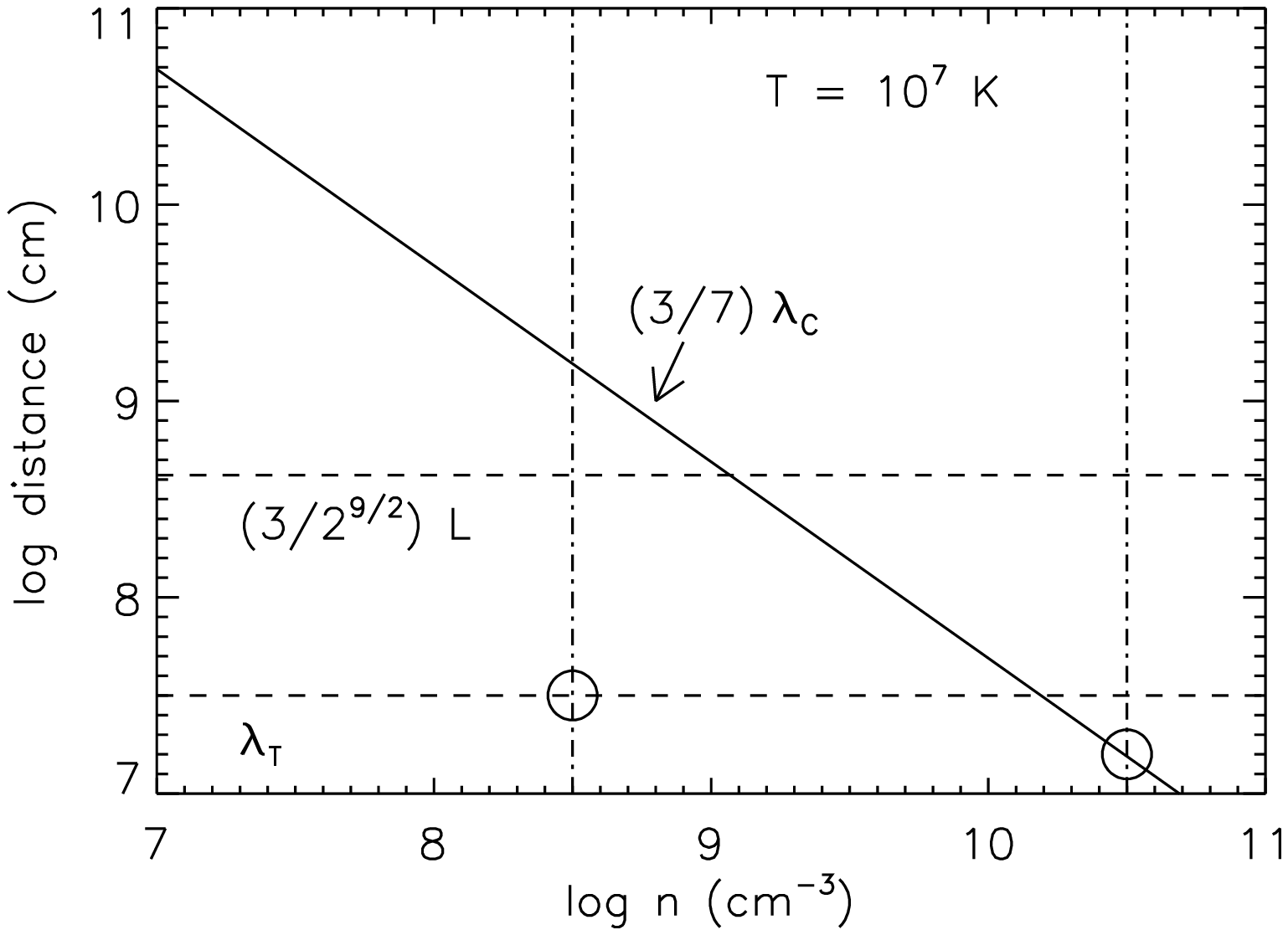}
	\includegraphics[width=0.49\linewidth]{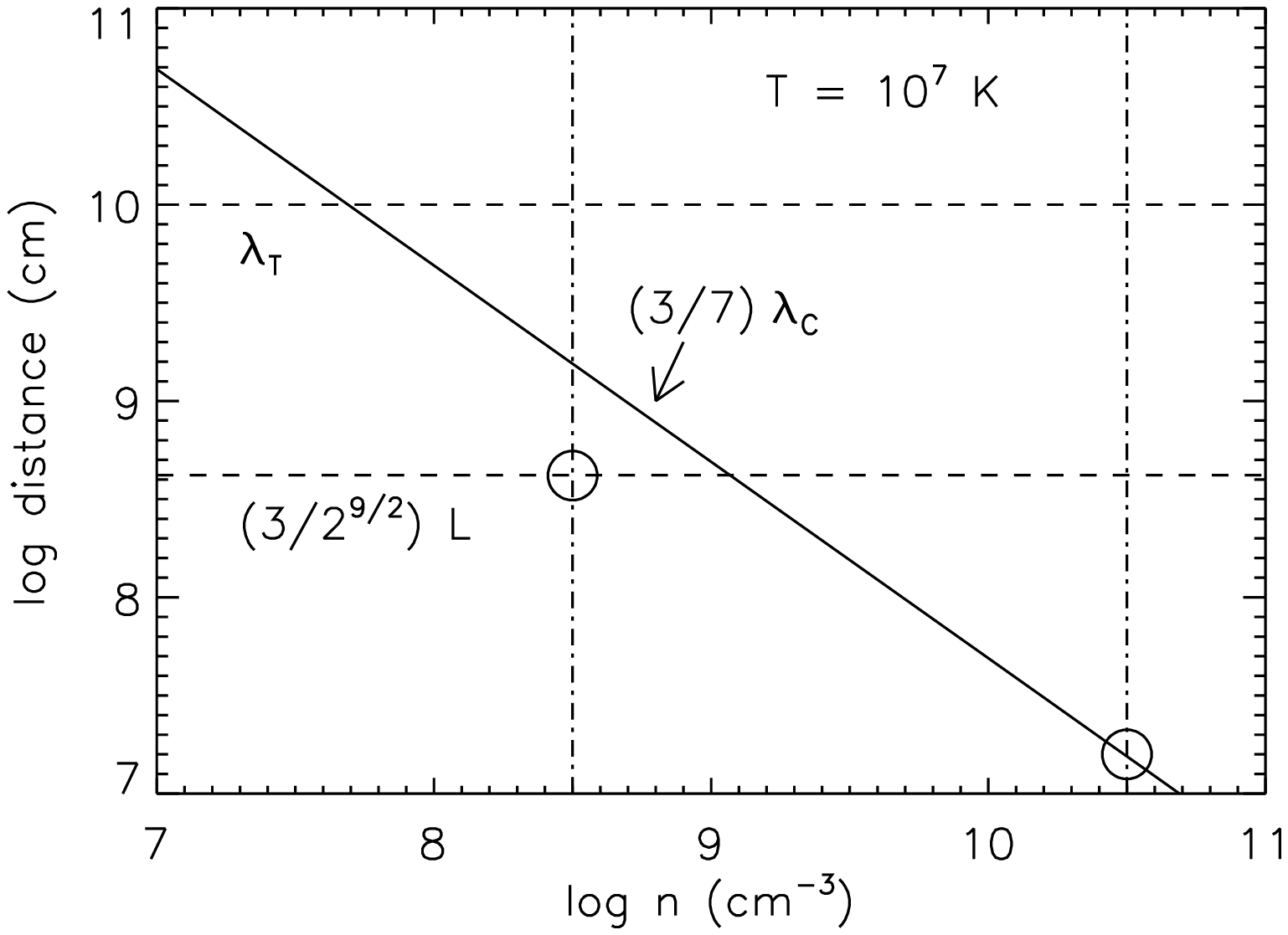}
	\caption{Values of pertinent spatial scales in different conduction regimes for $L=10^{9.5}$~cm and $T=10^7$~K. The diagonal line represents the quantity $(3/7)\lambda_C$, while the dashed horizontal lines represent the quantities $\lambda_T$ and $(3/2)^{9/2}L$, respectively. The dot-dashed vertical lines correspond to $n=10^{8.5}$~cm$^{-3}$ and $n=10^{10.5}$~cm$^{-3}$. The circles show where each of these vertical lines meet the lowest pertinent scale, and hence the relevant conduction and scaling-law regime. Thus, for $\lambda_T = 10^{7.5}$~cm, conduction is controlled by turbulence if $n=10^{8.5}$~cm$^{-3}$ and by collisions if $n=10^{10.5}$~cm$^{-3}$; whereas for $\lambda_T = 10^{10}$~cm, conduction is controlled by free-streaming if $n=10^{8.5}$~cm$^{-3}$ and by collisions if $n=10^{10.5}$~cm$^{-3}$.\label{scale-length-comparison-2D}}
\end{figure}

The results are as follows:

\begin{itemize}

\item {[$\lambda_T = 10^{7.5}$~cm, $L = 10^{9.5}$~cm.]} In this case, for the first set of parameters ($T = 10^{6.2}$~K, $n = 10^{10.5}$~cm$^{-3}$), the collision-related scale is the lowest and hence determines the rate of heat loss by conduction; in such a case the pertinent scaling laws are the \cite{1978ApJ...220..643R} scaling laws -- denoted by $[C]$ in Equations~(\ref{scaling-law-tp-all}) and~(\ref{scaling-law-ehp-all}). For the second set of parameters ($T = 10^{7.5}$~K, $n = 10^{9.0}$~cm$^{-3}$), the lowest scale length is the turbulence scale length $\lambda_T = 10^7$~cm (blue dot on red horizontal surface in left panel of Figure~\ref{scale-length-comparison}; in such a case the pertinent scaling laws are those denoted by $[T]$ in Equations~(\ref{scaling-law-tp-all}) and~(\ref{scaling-law-ehp-all}).
\item {[$\lambda_T = 10^{10}$~cm, $L = 10^{9.5}$~cm.]} For the first set of parameters ($T = 10^{6.2}$~K, $n = 10^{10.5}$~cm$^{-3}$), the collision-related scale is still the lowest; the pertinent scaling laws are still the \cite{1978ApJ...220..643R} scaling laws -- denoted by $[C]$ in Equations~(\ref{scaling-law-tp-all}) and~(\ref{scaling-law-ehp-all}). For the second set of parameters ($T = 10^{7.5}$~K, $n = 10^{9.0}$~cm$^{-3}$), the lowest scale length is now that related to the loop half-length $L$ (blue dot on green horizontal surface in right panel of Figure~\ref{scale-length-comparison}; in such a case the heat conduction is controlled by free-streaming and the pertinent scaling laws are now those denoted by $[S]$ in Equations~(\ref{scaling-law-tp-all}) and~(\ref{scaling-law-ehp-all})
\end{itemize}

\section{Discussion and Conclusions}\label{discussion}

This paper has extended the work of \cite{1978ApJ...220..643R} 
to include situations where the thermal conductive flux that redistributes heat 
within a coronal loop is governed by processes other than Coulomb collisions, specifically turbulent scattering and free-streaming.  Equations~(\ref{scaling-law-tp-all}) and~(\ref{scaling-law-ehp-all}) provide the pertinent scaling laws for peak temperature $T_{M}$, and volumetric heating 
rate $E_H$, respectively in terms of the loop pressure $p$ and half-length $L$.
It is notable that, because of the much weaker dependence of the thermal conduction coefficient $\kappa$ on temperature for the case of turbulent scattering by magnetic fluctuations, the characteristics of the loop in such a regime are governed not by the highest temperatures in the loop (as they are for both the collisional and free-streaming cases), but by conditions at the low-temperature (transition region) part of the loop.

Which of these scaling laws is appropriate in a particular environment depends on the ratios of the turbulent scale length $\lambda_T$ to the collisional mean free path $\lambda_C$ to the loop half-length $L$ (Equation~(\ref{length-comparison})). Figure~\ref{scale-length-comparison} illustrates examples where each process, 
and hence scaling law, dominates.

Given the likely role of turbulence in active region loops, 
particularly those associated with flaring activity \citep[][and references therein]{2018ApJ...852..127B}, and modern observations of the faint, hot component of emission in non-flaring active regions, considered a signature of impulsive heating \citep[e.g.,][]{2009ApJ...704L..58R,2009ApJ...698..756R,2009ApJ...693L.131S,2009ApJ...704..863S,2011ApJ...728...30T,2012A&A...544A.139M,2014ApJ...790..112B,2014A&A...564A...3P,2018ApJ...864....5M}, that allow more precise estimation of loop temperatures and densities \citep[and of the presence of turbulence;][]{2017PhRvL.118o5101K}, we encourage the comparison of observed loop parameters with these extended scaling laws, as a possible diagnostic of the physical conditions in active region coronal loops and hence of the energy required to create and sustain them.

\acknowledgements

SJB is grateful to the NSF for supporting this work through CAREER award AGS-1450230.  AGE and NB were supported by grant NNX17AI16G from NASA's Heliophysics Supporting Research program. EPK acknowledges the financial support from the STFC Consolidated Grant ST/P000533/1. We thank the referee for their expert review of the manuscript.

\bibliographystyle{aasjournal}
\bibliography{bib-scaling-laws}

\end{document}